\documentclass[a4paper,11pt]{article}

\usepackage{jheppub} 

\usepackage[T1]{fontenc} 

\newcommand{\slashed}{\slash \hspace{-0.19cm}}

\newcommand{\bfm}[1]{\mbox{\boldmath$#1$}}

\title{\boldmath Light quark mediated   Higgs boson production
in association with a jet  at the next-to-next-leading order and beyond
}

\author[a,b]{Tao Liu,}
\author[c]{Alexander A. Penin,}
\author[c]{Abdur Rehman}

\preprint{ALBERTA-THY-2-24}

\affiliation[a]{Institute of High Energy Physics,  Chinese Academy of Sciences, Beijing 100049, China}
\affiliation[b]{University of Chinese Academy of Sciences, Beijing 100049, China}
\affiliation[c]{Department of Physics, University of Alberta,
Edmonton AB T6G 2J1, Canada}

\emailAdd{liutao86@ihep.ac.cn}
\emailAdd{penin@ualberta.ca}
\emailAdd{rehman3@ualberta.ca}

\abstract{
We study the light quark effect on the  Higgs boson production
in association with a jet at the LHC in the intermediate
transverse momentum region between the quark and the Higgs
boson mass scales. Though the effect  is suppressed  by the small
Yukawa coupling, it is enhanced by large logarithms of the
quark mass ratio to the Higgs boson mass or transverse
momentum. Following  a remarkable success  of the logarithmic
expansion \cite{Anastasiou:2020vkr} for  the prediction of the
next-to-next-to-leading bottom quark contribution to the total
cross section of the Higgs boson production   we extend the
analysis to its kinematical  distributions. A new factorization
formula is derived for the light quark mediated $gg\to Hg$
amplitudes and  the differential cross section of the process
is computed  in the logarithmic approximation, which is used
for an estimate of the bottom quark effect  at the
next-to-next-to-leading order.}

\begin{document}
\maketitle
\flushbottom

\section{Introduction}
\label{sec::int}

Determination of the Higgs boson properties  with high
precision is a major part of the Large Hadron Collider  physics
program \cite{ATLAS:2016neq,Cepeda:2019klc}. A significant
deviation of the measured couplings from the theory predictions
may indicate a signal of the physics beyond the standard model.
The  transverse momentum distribution of the Higgs boson
produced in association  with a jet is an important kinematical
observable sensitive to its Yukawa couplings and  can be used
to test different  extensions of  the standard model in
different  regions of the transverse momentum
\cite{Arnesen:2008fb,Bagnaschi:2011tu,Dawson:2014ora,Grazzini:2016paz}.
The dominant contribution to the process cross section is given
by the top quark loop mediated gluon fusion. It is known through
the next-to-next-to-leading order (NNLO) in the strong coupling
constant in the heavy top quark effective theory
\cite{Chen:2014gva,Boughezal:2015dra,Boughezal:2015aha,Caola:2015wna,Chen:2016zka}
and recently has been evaluated to next-to-leading order (NLO)
with the full top quark mass dependence \cite{Jones:2018hbb}.
The contribution of the light quarks to the process is strongly
suppressed by their  masses. This suppression is partially
compensated by the enhancement of the leading order amplitude
by the second power of the large logarithm $\ln(m_H^2/m_q^2)$
of the Higgs boson to the light quark mass ratio
\cite{Baur:1989cm}, and the bottom quark effect can be as large
as $5\%$ \cite{Anastasiou:2016cez}. Moreover, the contribution
can be distinguished by its  transverse momentum dependence for
$p_\perp \ll m_H$, where the cross section is large. This makes
the  differential distribution  sensitive to the light quark
Yukawa couplings \cite{Bishara:2016jga,Soreq:2016rae}. Thus, it
is important to get an accurate prediction for the differential
cross section in perturbative QCD. The NLO result in the small
quark mass limit has been derived in
\cite{Lindert:2017pky,Bonciani:2022jmb}. The higher order
corrections may also be important  due to the presence of the
double-logarithmic terms with the second power of the large
logarithm per each power of the strong coupling constant. The
resummation of the logarithmically enhanced terms poses a
serious theoretical challenge for the effective field theory
methods
\cite{Mantler:2012bj,Grazzini:2013mca,Banfi:2013eda,Caola:2018zye}
due to their non-standard  origin
\cite{Kotsky:1997rq,Penin:2014msa,Liu:2017axv}. In contrast to
the standard Sudakov case
\cite{Sudakov:1954sw,Frenkel:1976bj,Mueller:1979ih,Collins:1980ih,Sen:1981sd,
Sterman:1986aj},  the logarithmic correction to the
mass-suppressed amplitudes are determined by the  eikonal color
charge nonconservation  in the process of the soft quark
exchange \cite{Liu:2017vkm}. The comprehensive analysis of this
phenomena in the leading (double) logarithmic approximation has
been performed for the three-point amplitudes in QED and QCD
\cite{Liu:2018czl,Liu:2021chn}. For the four-point  amplitudes
the abelian part of the all-order result is available for the
$gg\to Hg$ production \cite{Melnikov:2016emg} and only the
two-loop result for Bhabha scattering
\cite{Penin:2016wiw,Delto:2023kqv}.\footnote{The all-order
leading logarithmic result is known for the subleading power
contribution in the Regge limit \cite{Penin:2019xql}.} The
resummation of the subleading logarithms has been  also
completed  for the  $gg\to H$ amplitudes
\cite{Anastasiou:2020vkr,Liu:2020wbn,Liu:2022ajh}. The
logarithmic expansion turned out to be remarkably successful
for the estimate of the bottom quark contribution to the total
cross section of the  Higgs  boson production in the NNLO. The
next-to-leading-logarithmic (NLL) approximation for this
quantity gives  $2.18\pm 0.20~pb$ \cite{Anastasiou:2020vkr},
while the recently computed full result reads
$1.99(1)\,{}^{+0.30}_{-0.15}~pb$
\cite{Czakon:2023kqm}.\footnote{The discrepancy in the
next-to-next-to-leading order terms quoted in
\cite{Czakon:2023kqm} is mainly due to the different bottom
quark mass renormalization scheme adopted in the paper.} Hence,
the generalization of the analysis to the case of the final
state jet seems quite important both for general theory of
renormalization group  at subleading power and from purely
phenomenological point of view.

In this paper we extend the abelian result
\cite{Melnikov:2016emg} to the full QCD description of the
light quark contribution to the differential cross section of
the  Higgs boson production in association with a jet  for the
intermediate values of transverse momentum. This requires a
deeper understanding of the origin of the non-Sudakov
logarithms and a systematic derivation of the factorization
formula.  The paper is organized  as follows. In the next
section we introduce our notations. In Sect.~\ref{sec::3} we
start with the factorization analysis of the double-logarithmic
contribution to the leading order one-loop amplitude, then
derive the general all-order resummation formula and compare it
to the available two-loop expressions \cite{Melnikov:2016qoc}.
In Sect.~\ref{sec::4} we apply this result to the analysis of
the $pp\to Hj+X$ differential cross section, which determines
the NNLO bottom quark contribution in terms of the known
$K$-factors obtained in the heavy quark limit.
Sect.~\ref{sec::conc} is our conclusion.

\section{Setup and notations}
\label{sec::2}

We consider a light quark  mediated   Higgs boson production
in the process
\begin{equation}
g(p_1)+g(p_2)\to g(p_3)+H(p_H)
\label{eq::process}
\end{equation}
for the intermediate values of the  transverse momentum
\begin{equation}
m_q^2 \ll  p_\perp^2  \ll s,m_H^2\,,
\label{eq::kincond}
\end{equation}
where $p_\perp^2=ut/s$,  $s = (p_1+p_2)^2, t = (p_1 - p_3)^2,
u=(p_2-p_3)^2$. We focus on the near-threshold production for
partonic center-of-mass energy  $s\approx m_H^2$ so that the
additional soft gluon emission energy  is small compared to the
leading jet. This, in particular, means that the  transverse
momenta of the Higgs boson and the jet are approximately equal
in magnitude. The above kinematical region gives the dominant
contribution to the hadronic cross section.  For $p_\perp^2\ll
s$ the process is described by four helicity amplitudes
$M_{\lambda_1\lambda_2\lambda_3}$  with the same helicity of
colliding gluons $\lambda_1=\lambda_2$, which are pairwise
connected by spatial inversion.  We take $M_{++\pm}$ as the two
independent amplitudes and only keep the leading terms which
grow as $1/p_\perp$ with decreasing transverse momentum.   In
the spinor notations the corresponding expressions read
\begin{eqnarray}
M_{+++}&=&-\sqrt{2}f^{a_1a_2a_3}{g_s\over v}\frac{\alpha_s}{4\pi}
\frac{\langle12\rangle^2}{[12]\langle23\rangle\langle13\rangle }
Z_{3g}\sum_q A_{+++}^{(q)}\,,
\\
M_{++-}&=&-\sqrt{2} f^{a_1a_2a_3}{g_s\over v}\frac{\alpha_s}{4\pi}
\frac{\langle12\rangle }{[23][13]}
Z_{3g}\sum_q A_{++-}^{(q)}\,,
\end{eqnarray}
where $\alpha_s=g_s^2/(4\pi)$ is the strong coupling constant,
$v$ is the Higgs field vacuum expectation value,
$f^{a_1a_2a_3}$ is the $SU(3)$ structure constant, the sum goes
over the quark flavors, $Z_{3g}$ is the Sudakov factor
incorporating the soft and collinear divergences of the
amplitudes, and  $A_{++-}^{(q)}$ are the infrared finite form
factors. Let us first discuss the leading contribution of the
top quark which can be expanded in the inverse  powers of
$m_t$. The perturbative series for the form factors then take
the following form
\begin{equation}
A_{++\pm}^{(t)}=C_t\sum_{n=0}
\left({\alpha_s\over 2\pi}\right)^n
\tilde A^{(n)}_{++\pm}+{\cal O}(m_t^{-2})\,,
\label{eq::topamp}
\end{equation}
where the Wilson coefficient $C_t=1+{11\over
4}{\alpha_s\over\pi}+\ldots$ is known through $\alpha_s^4$
\cite{Schroder:2005hy,Chetyrkin:2005ia}, $\tilde
A^{(0)}_{++\pm}=\pm {2/3}$, and $\tilde A^{(n)}_{++\pm}$ stands
for the $n$-loop correction computed  in the heavy top
effective theory, which is   available up
to $n=2$ \cite{Gehrmann:2011aa}. For light  quarks the
amplitudes can be expanded in powers of $m_q$ and the
perturbative series for the form factors reads
\begin{equation}
A_{++\pm}^{(q)}={m_q^2\over s}\sum_{n=0}
\left({\alpha_s\over 2\pi}\right)^n
A^{(n+1)}_{++\pm}+{\cal O}(m_q^4)\,,
\label{eq::qamp}
\end{equation}
where $A^{(n)}_{++\pm}$ stands for the $n$-loop QCD
contribution also available up to $n=2$
\cite{Melnikov:2016qoc}. These coefficients include the terms
enhanced by the large logarithms of the  scale ratios
$p_\perp^2/m_q^2$, $m_H^2/m_q^2$ or  $t/u$, which becomes
relevant at large rapidity.  The goal of this work is to
compute the coefficients $A^{(n)}_{++\pm}$ in the leading
(double) logarithmic approximation, {\it i.e.} keeping  the
highest power  of the large logarithms equal to $2n$, for all
$n$.

In the leading  logarithmic approximation the  Sudakov factor
can be written as follows $Z^{LL}_{3g}=e^{{\alpha_s\over
2\pi}I^{(1)}}$ where $I^{(1)}$ is an analog of the one-loop
Catani's operator \cite{Catani:1998bh} with the  trivial
structure in the color space. It is a function of the
Mandelstam variables and is scheme dependent. To make the
factorization of the Sudakov and non-Sudakov logarithms
explicit it is convenient to define this function in such a way
that it incorporates the nonsingular double logarithmic
dependence of  the  one-loop corrections to the effective
theory amplitudes with the local $ggH$ interaction  on the
kinematical invariants. By using the result
\cite{Schmidt:1997wr} it is straightforward to  get the leading
singularity subtraction operator in the ``physical'' scheme
\begin{equation}
I^{(1)}_{\rm ph}=-{C_A\over 2\epsilon^2}
\left[2\left({-s\over \mu^2}\right)^{-\epsilon}
+\left({-tu\over s\mu^2}\right)^{-\epsilon}\right],
\label{eq::I1ph}
\end{equation}
where $\epsilon=2-d/2$ is the parameter of dimensional
regularization and $C_A=N_c$ for the $SU(N_c)$ color gauge
group. Note that the canonical  symmetric form of the
subtraction operator has been used in \cite{Melnikov:2016qoc},
which differs from Eq.~(\ref{eq::I1ph}) by a non-singular
double logarithmic terms, as discussed in Sect.~\ref{sec::3.2}.

\section{Factorization and resummation of the leading logarithms}
\label{sec::3}

\subsection{Factorization  of the  one-loop amplitudes}
\label{sec::3.1}

\begin{figure}[t]
\begin{center}
\begin{tabular}{ccc}
\hspace{0mm}\raisebox{23.0mm}{\tiny $1$}\hspace{-2.5mm}
\raisebox{-.5mm}{\tiny   $2$}\hspace{12.mm}
\raisebox{30.0mm}{\tiny   $3$}\hspace{-13.mm}
\includegraphics[width=3cm]{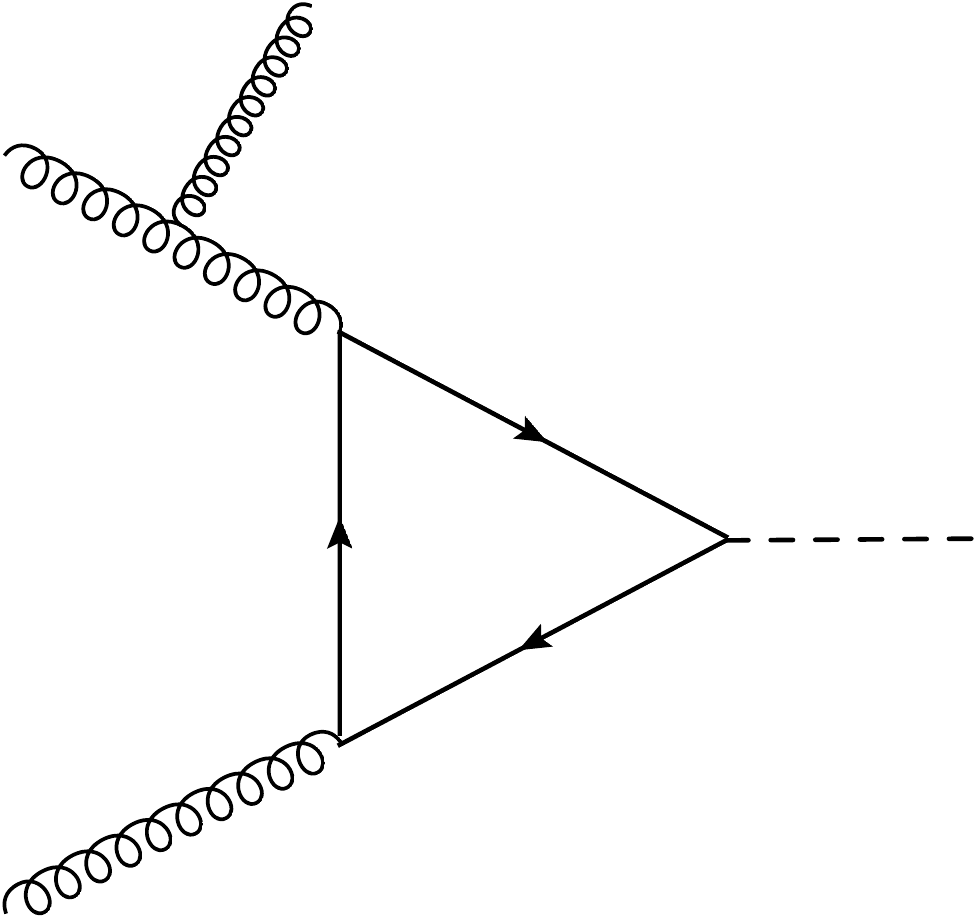}\hspace*{10.mm}&
\includegraphics[width=3cm]{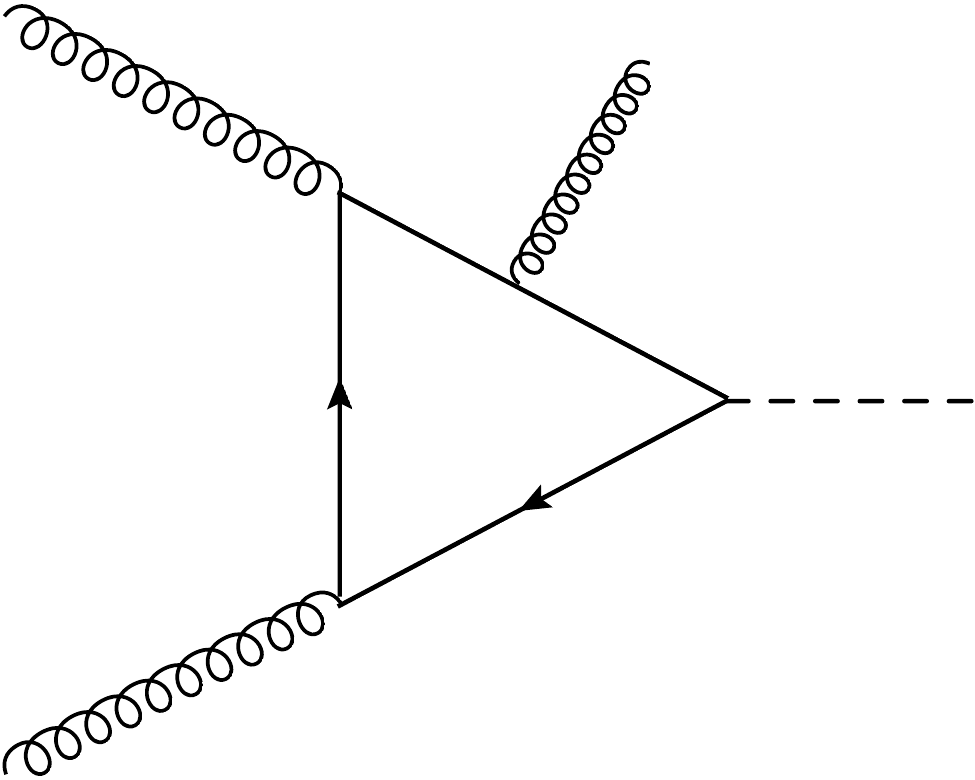}\hspace*{10.mm}&
\includegraphics[width=3cm]{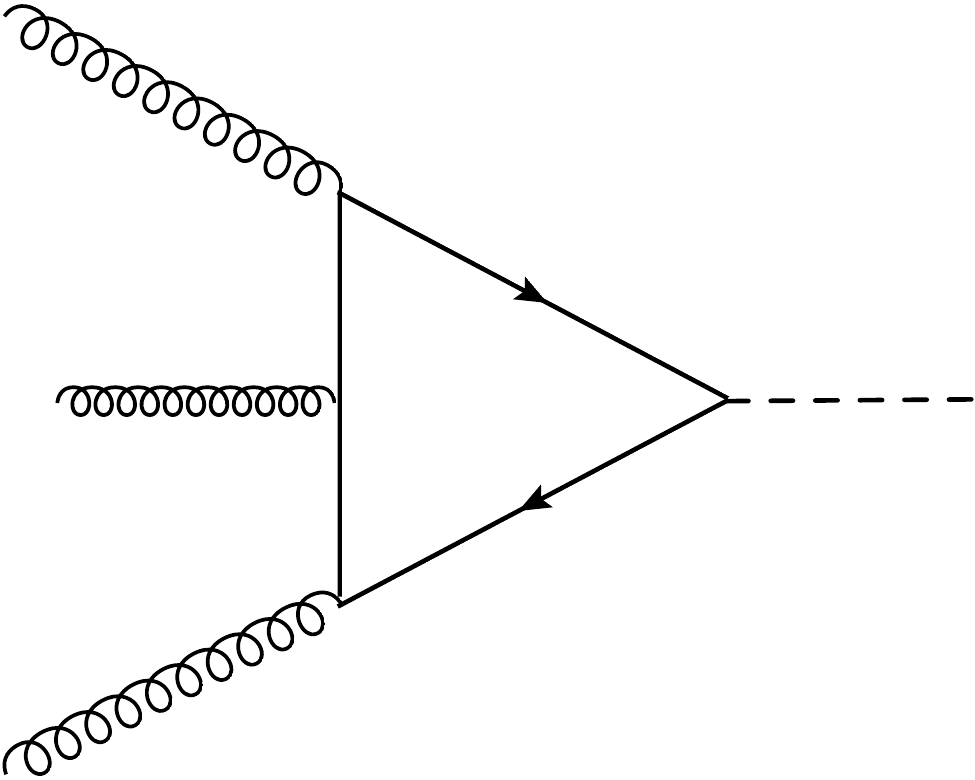}\\
&&\\
(a)\hspace*{14.mm}&(b)\hspace*{16.mm}&(c)\hspace*{8.mm}
\end{tabular}
\end{center}
\caption{\label{fig::1}
One-loop diagrams representing the leading order light quark
contribution to $gg \to Hg$ process. Wide and narrow gluon
lines correspond to the  initial and  final gluon states,
respectively. Symmetric diagrams corresponding to the opposite
direction of the quark flow and to the soft emissions off  the
opposite gluon/quark line are not shown. In the first diagram
the gluons  are explicitly enumerated according to their momentum index.}
\end{figure}

A detailed discussion of the evaluation of the leading order
one-loop amplitude in the double-logarithmic approximation can
be found in \cite{Melnikov:2016emg}. It, however, does not
reveal a deeper factorization structure, which is crucial for
the full QCD analysis and is discussed in this section.
Following \cite{Melnikov:2016emg} we choose the gluon
polarization vectors $\epsilon_i$ satisfying the gauge
conditions
\begin{equation}
\epsilon_i p_i=0\quad (i=1,2,3)\,,\qquad
\epsilon_1  p_2=0\,,\qquad
\epsilon_2 p_1 = 0\,,\qquad
\epsilon_3  p_2=0\,,
\label{eq:gauge}
\end{equation}
so that the jet cannot be emitted off the gluon or quark line
carrying the large light-cone  momentum $p_2$. It is convenient
to choose a frame where the light-cone components
${p_1}_\perp={p_2}_\perp=p_1^+=p_2^-=\epsilon_3^-=0$. Then, up
to the direction of the closed  quark line the leading order
process is described by the three diagrams in
Fig.~\ref{fig::1}. The diagrams Fig.~\ref{fig::1}(a,b) have the
standard structure of soft emission from a highly energetic
eikonal line.  In the diagram  Fig.~\ref{fig::1}(c), however,
the jet is emitted from the soft quark line which does not
carry the large momenta $p_{1,2}$. We focus on the specific
structure of the corresponding double logarithmic contribution.
Due to the helicity conservation in the small mass limit and
the helicity flipping scalar Yukawa interaction, the light
quark contribution is mass suppressed.  The mass-suppressed
double-logarithmic corrections are known to be generated by the
on-shell soft quark contribution. Thus the quark propagator in
the expression for the Feynman diagrams in Fig.~\ref{fig::1}
can be approximated as follows
\begin{equation}
S(l)\equiv {1\over \slashed{l}-m_q+i0}\to-im_q\pi
\delta(l^2 - m_q^2)\,,
\label{eq::softprop}
\end{equation}
where $l$ is the soft loop momentum and the mass factor
accounts for the required helicity flip on the virtual quark
line.

Each of the two t-channel quark propagators in
Fig.~\ref{fig::1}(c) can go on-shell which defines two soft
virtual momentum regions. Let us consider the one with the
upper on-shell propagator so that the jet momentum $p_3$ flows
through the lower one. After omitting irrelevant terms, the
off-shell quark propagators become
\begin{eqnarray}
&&S(p_1+l)\to \frac{\slashed{p_1}}{ 2p_1 l}\,,
\label{eq::prop1}
\\
&&S(p_3+l) \to
\frac{\slashed{p_3}+\slashed{l}}{2p_3 l}\,,
\label{eq::prop2}
\\
&&S(p_3-p_2+l) \to  \frac{\slashed{p_2}}{2p_2p_3+2p_2l}\,.
\label{eq::prop3}
\end{eqnarray}
Then  the integral develops the double logarithmic scaling
either with the first term in the numerator of
Eq.~(\ref{eq::prop2}) for $|p_2l|< |p_2p_3|$ or with the second
term for $|p_2p_3|< |p_2l|$.  Thus we can write
\begin{eqnarray}
S(p_3+l)\slashed{\epsilon_2} S(p_3-p_2+l)&\to&
\theta\left(|p_2p_3|-|p_2l|\right)
{\slashed{p_3}\over 2p_3l}
\slashed{\epsilon_2} {\slashed{p_2}\over
2p_2p_3}
\nonumber\\
&&+\theta(|p_2l|-|p_2p_3|){\slashed{l}\over 2p_3l}\slashed{\epsilon_2}
{\slashed{p_2}\over 2p_2l}\,.
\label{eq::propdec}
\end{eqnarray}
This decomposition, however, does not reflect the factorization
structure in the effective theory terms and we rewrite it as
follows
\begin{eqnarray}
S(p_3+l) \slashed{\epsilon_2} S(p_3-p_2+l)&\to&\theta\left(|p_2p_3|-|p_2l|\right)
\left({\slashed{p_3}\over 2p_3l} \slashed{\epsilon_2 }{\slashed{p_2}\over
2p_2p_3}-{\slashed{l}\over 2p_3l}\slashed{\epsilon_2}{\slashed{p_2}\over
2p_2l}\right)
\nonumber\\
&&+{\slashed{l}\over 2p_3l}\slashed{\epsilon_2}
{\slashed{p_2}\over 2p_2l}\,,
\label{eq::propdeceff}
\end{eqnarray}
where the integration region is extended in the second term and
its variation is subtracted from the first term.

\begin{figure}[t]
\begin{center}
\begin{tabular}{ccc}
\includegraphics[width=3cm]{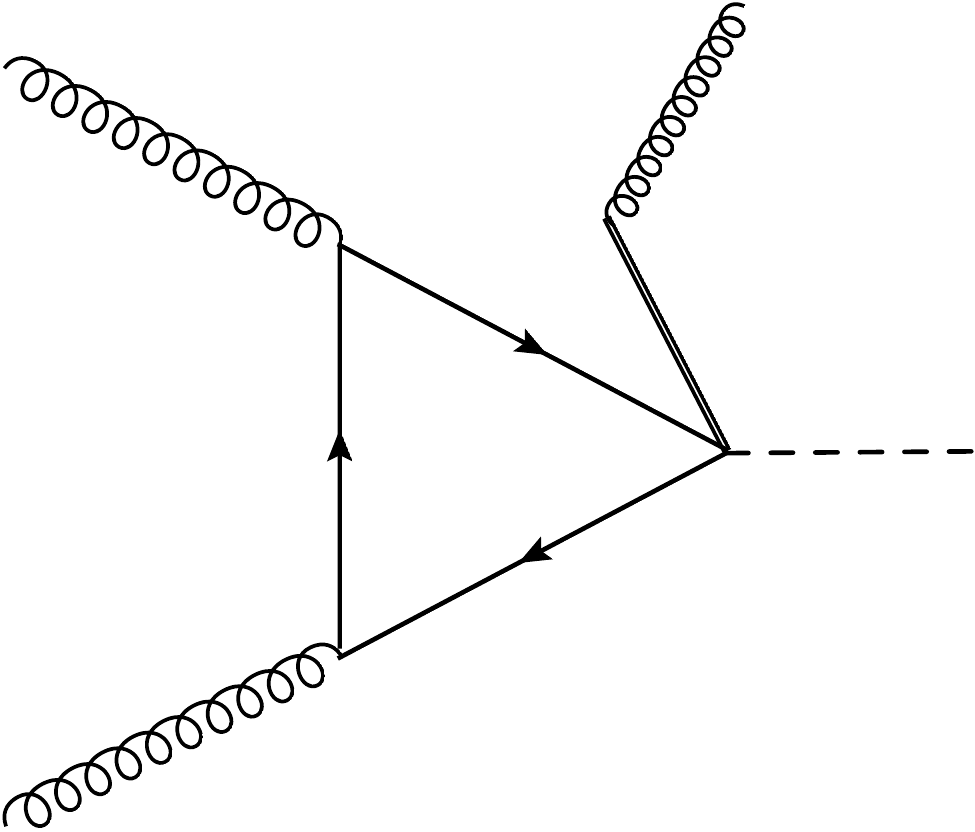}\hspace*{10.mm}&
\includegraphics[width=3cm]{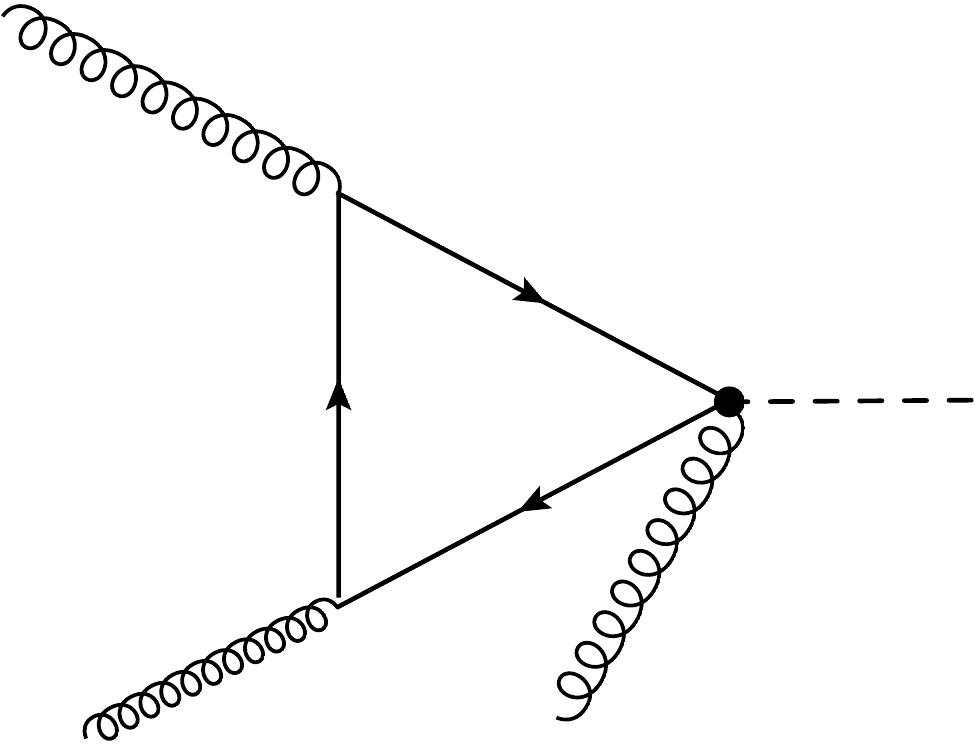}\hspace*{10.mm}&
\includegraphics[width=3cm]{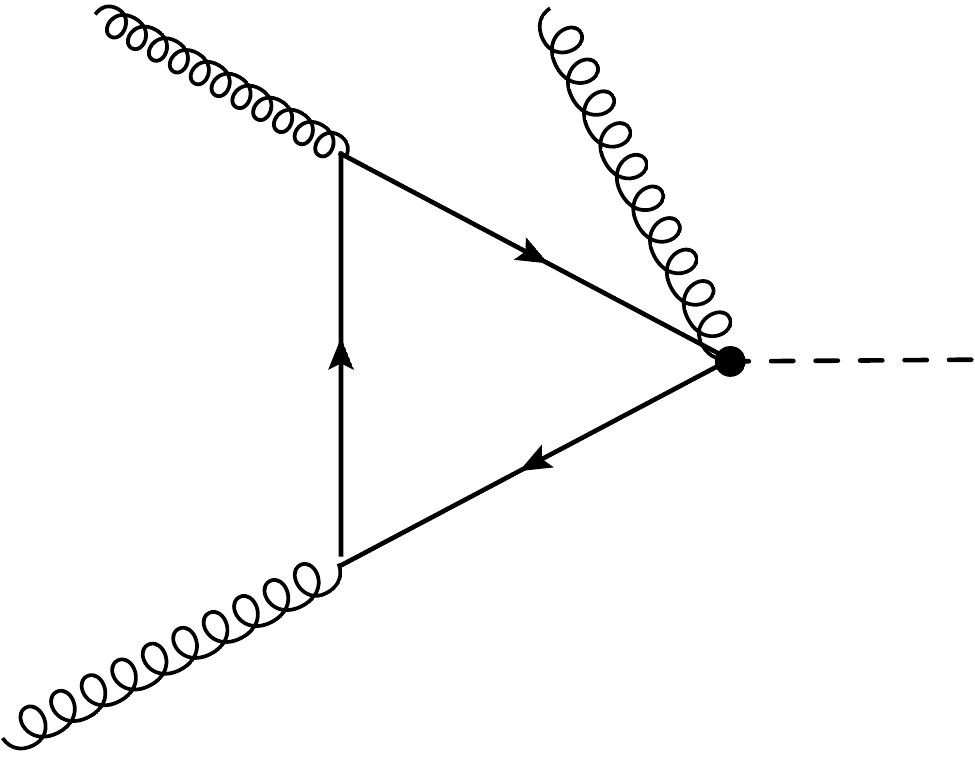}\\
&&\\
(a)\hspace*{14.mm}&(b)\hspace*{16.mm}&(c)\hspace*{8.mm}
\end{tabular}
\end{center}
\caption{\label{fig::2}
The effective representation of the sum of the one-loop
diagrams in Fig.~\ref{fig::1} after decomposition into the {\it
dipole} and {\it symmetric} structures. The double line
represents the eikonal line defined by the momentum $p_1$, as
explained in the text. The black circle corresponds to the
effective vertex.}
\end{figure}

Let us first consider the contribution   proportional to the
theta-function in  Eq.~(\ref{eq::propdeceff}). It includes two
distinct structures corresponding  to two terms in the
brackets. For the second term in  the logarithmic integration
region ${\slashed{\,l}\over 2p_3l}$ can be replaced by
${\slashed{p_1}\over 2p_3p_1}$, which results in the amplitude
proportional to  $\epsilon_3p_1\over p_3p_1$ characteristic to
the color-dipole  emission from the eikonal line of the
momentum $p_1$, and gives   $A_{+++} = - A_{++-}$. Moreover,
the only dependence  on the momentum $p_3$ is through the
argument of the theta-function and the integrand coincides with
the one of the one-loop $ggH$ amplitude. Thus this {\it soft
dipole} contribution reduces to the effective diagram
Fig.~\ref{fig::2}(a) with the decoupled jet and a constraint
$|p_2l|<|p_2p_3|$ on the loop momentum. The double line in this
diagram stands for the eikonal (Wilson) line defined by the
light-like  momentum $p_1$, with the propagator
$-p_1/(2p_3p_1)$. The calculation of this contribution is
straightforward by using the classical Sudakov technique  which
converts the integral over $l$ into the integral over the
logarithmic Sudakov variables $\eta=\ln (|lp_3|/|p_1p_3|)/L$
and $\xi=\ln (|lp_1|/|p_2p_1|)/L$ (see \cite{Melnikov:2016emg}
for the details)\footnote{In \cite{Melnikov:2016emg} the second
logarithmic variable is defined as $\xi=\ln
(|lp_1|/|p_3p_1|)/L$.}
\begin{equation}
\left[A^{(1)}_{++ \pm}\right]_{\rm s.d.} =\mp L^2
\int_{\tau_t-\tau}^{\tau_t}{\rm d}\eta
\int_{1-\tau_t}^{1-\eta}{\rm d}\xi
=\mp{L^2}{\tau^2\over 2}\,,
\label{eq::softdip}
\end{equation}
where $L=\ln (s/m_q^2)$,
$\tau_t=\ln\left({|t|/m_q^2}\right)/L$, and
$\tau=\ln\left({p_\perp^2/m_q^2}\right)/L$. Note that a dipole
contribution  from the  virtual momentum configuration where
the lower t-channel propagator goes on-shell and  the jet
momentum $p_3$ flows through the upper one is proportional to
$\epsilon_3p_2$ and vanishes for our gauge condition.

The first term in the brackets in Eq.~(\ref{eq::propdeceff})
proportional to $p_3$ results in a new symmetric tensor
structure  described by a gauge invariant operator
$G_{\mu\nu}^aG^{b}_{\nu\lambda}G_{\lambda\mu}^cf^{abc}$, which
does not contribute   to the all-plus  helicity amplitude. It
can be represented by the effective theory diagram
Fig.~\ref{fig::2}(a) with the propagator $S(p_2-p_3+l)$
collapsed to an  effective local vertex.  Due to its symmetry
and explicit gauge invariance  this structure does get the
equal contribution from the crossing-symmetric configuration
with the opposite direction of the jet momentum flow
represented by the effective diagram in Fig.~\ref{fig::2}(b).
The sum gives the {\it symmetric} contribution
\begin{equation}
\left[A^{(1)}_{++-}\right]_{\rm sym.} =- L^2\left(
\int_{\tau_t-\tau}^{\tau_t}{\rm d}\eta\int_{1-\tau_t}^{1-\eta}
{\rm d}\xi+(\tau_t\leftrightarrow \tau_u)\right)
=-{L^2}{\tau^2}\,,
\label{eq::sym}
\end{equation}
where $\tau_u=\ln\left({|u|/m_q^2}\right)/L$.  The total
contribution of the theta-function term  then reads
\begin{equation}
\left[A^{(1)}_{++\pm}\right]_{\rm s.d.}
+\left[A^{(1)}_{++\pm}\right]_{\rm sym.} =-{\tau^2\over 2}L^2\,.
\label{eq::restheta}
\end{equation}
It is helicity independent and vanishes at the boundary of the
transverse momentum region $p_\perp \to m_q$ where $\tau\to 0$.

\begin{figure}[t]
\begin{center}
\begin{tabular}{cccc}
\includegraphics[width=3cm]{fig3.pdf}\hspace*{0.0mm}
\raisebox{10.5mm}{$\bfm{\to} ~{1\over 2}$}
\hspace*{00mm}&
\hspace*{-5mm}
\includegraphics[width=3cm]{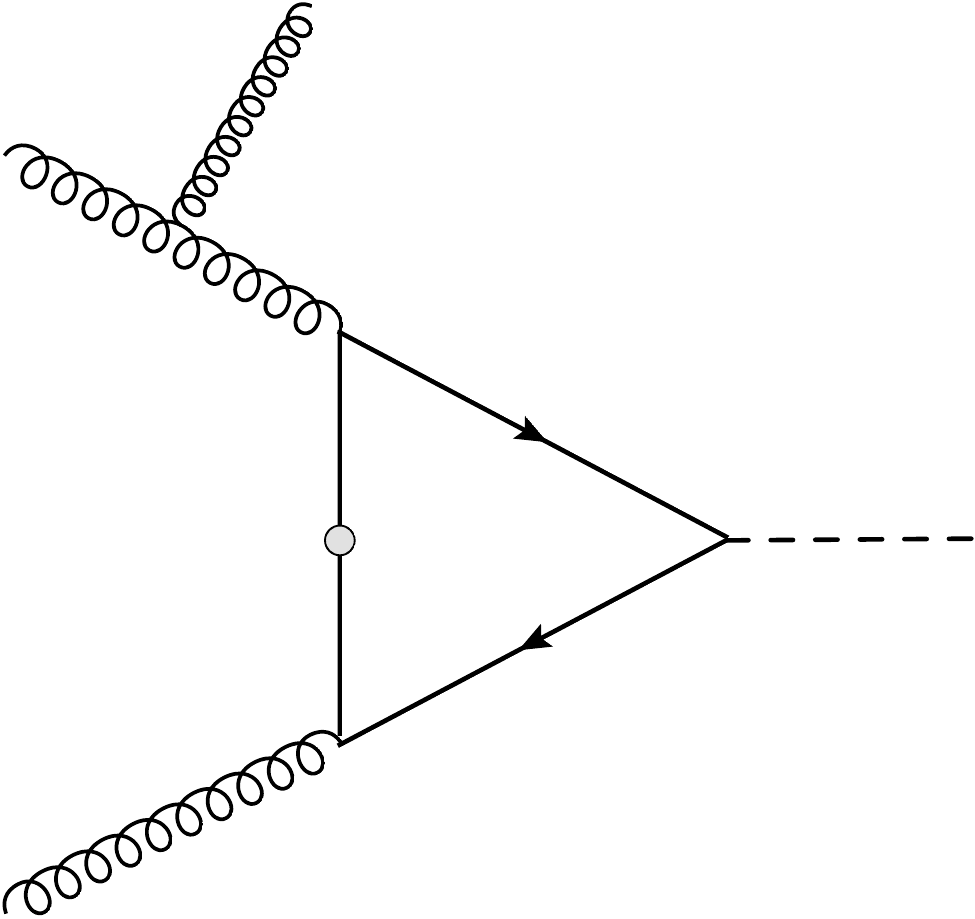}\hspace*{0.0mm}
\raisebox{10.5mm}{$\bfm{\to}~{1\over 2}$}
\hspace*{00mm}&
\hspace*{-5mm}
\includegraphics[width=3cm]{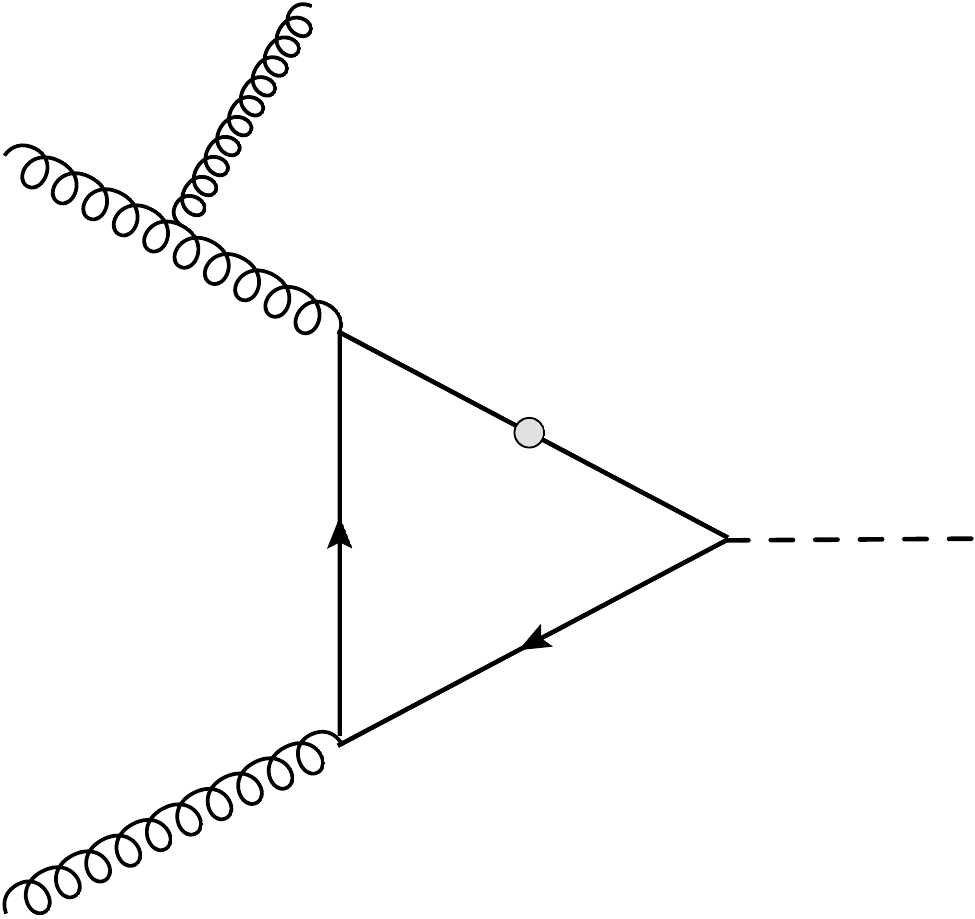}\hspace*{0.0mm}
\raisebox{10.5mm}{$\bfm{\to}$}
\hspace*{00mm}&
\hspace*{-7mm}
\includegraphics[width=3cm]{fig2.pdf}
\hspace*{00mm}\\
&&&\\
(a)&(b)&(c)&(d)
\end{tabular}
\end{center}
\caption{\label{fig::3}
Diagrammatic representation of the sequence of the Ward
identities and momentum shifts  which relates the unconstrained
dipole contribution of the diagram Fig.~\ref{fig::1}(c) to the
diagram Fig.~\ref{fig::1}(b), as explained in the text. The
empty circle marks  the subtracted quark propagator.}
\end{figure}

To account for the unconstrained dipole contribution  of the
second line in Eq.~(\ref{eq::propdeceff}) we use the method of
\cite{Liu:2018czl} and move the jet vertex from the soft quark
propagator to the eikonal line, Fig~\ref{fig::3}. This can be
done  in three steps. First we  transform the soft line as
follows
\begin{equation}
S(l) \slashed{\epsilon_3} S(l+p_3)=S(l)
\slashed{\epsilon_3} S(l+p_3^+)+\ldots=
{\epsilon_3 p_1\over p_3p_1}
\left(S(l)- S(l+p_3^+)\right)+\ldots\,,
\label{eq::Ward}
\end{equation}
where we used the fact that $\epsilon_3^-=0$, and  the omitted
terms do not contribute to the leading logarithms. This
transformation, being a particular realization of the
Slavnov-Ward identities,  is graphically represented by the
diagram Fig.~\ref{fig::3}(b), where the empty circle on the
quark propagator represents the replacement $S(l)\to
S(l)-S(l+p_3^+)$  and the factor $1/2$ comes from the color
weight of the diagrams. By the momentum shift $l\to l-p_3^+$ in
the second term of the above expression the crossed circle can
be moved to the upper eikonal quark line  which becomes
$S(p_2+l)-S(p_2+l-p_3^+)$, Fig.~\ref{fig::3}(c). The opposite
eikonal line is not sensitive to this shift since $p_2^-=0$. On
the final step we use the  ``inverted identity'' on the upper
eikonal quark line
\begin{equation}
{\epsilon_3 p_1\over p_3p_1}\left(S(p_2+l)- S(p_2+l-p_3^+)\right)
= -S(p_2+l)\slashed{\epsilon_3}S(p_2+l-p_3) +\ldots\,,
\label{eq::Wardinv}
\end{equation}
to  transform the diagram Fig.~\ref{fig::2}(c) into
Fig.~\ref{fig::2}(d) equal to Fig.~\ref{fig::1}(b). Now we can
combine this  dipole part of the diagram Fig.~\ref{fig::1}(c)
with the diagrams  Fig.~\ref{fig::1}(a,b) to get  the
standard eikonal factorization Fig.~\ref{fig::1}(a)$+$2
Fig.~\ref{fig::1}(b)$=$ Fig.~\ref{fig::2}(a) where the factor 2
accounts for the difference in the color weights. In
Fig.~\ref{fig::2}(a) the jet is emitted by the eikonal Wilson
line of momentum $p_1$ and is completely decoupled from the
quark loop without the  constraint on the loop momentum present
in the soft dipole contribution. Thus this {\it eikonal dipole}
contribution reduces to the one-loop $g(p_1)g(p_2)H$ form factor
\begin{equation}
\left[A^{(1)}_{++\pm}\right]_{\rm e.d.} =\pm 2 L^2
\int_{0}^{1}{\rm d}\eta\int_0^{1-\eta}{\rm d}\xi=
\pm L^2\,,
\label{eq::reseik}
\end{equation}
and is independent of the jet kinematics. By adding up
Eqs.~(\ref{eq::restheta}) and (\ref{eq::reseik}) we get
\begin{equation}
A^{(1)}_{+++} = L^2\left(1-{\tau^2\over 2}\right),
\qquad  A^{(1)}_{++-} = -L^2\left(1+{\tau^2\over 2}\right),
\label{eq::res2loop}
\end{equation}
in agreement with \cite{Baur:1989cm}.

The main result of this section is the decomposition of the
one-loop amplitudes into the eikonal dipole, soft dipole,  and
symmetric parts. In the first one the jet emission factors out
from the quark loop, as it  happens for $p_\perp\ll m_q$, {\it
i.e.} in the heavy quark limit with the effective
$g(p_1)g(p_2)H$ vertex. The last two describe the deviation
from this factorization, accommodate all the dependence on the
jet kinematical variables, and vanish when  $p_\perp\to m_q$.
The soft dipole contribution reduces to the quark-loop mediated
$g(p_1)g(p_2)H$ form factor with a new {\it ultraviolet}
constraint on the double logarithmic region of loop momentum
depending on jet kinematics. The symmetric contribution  is
characterized by a new local effective vertex involving  one of
the initial state gluons and reduces to the  quark-loop
mediated $g(p_3)g(p_{1,2})H$ form factors with one of the gluons
being the jet and with the same constraint on the double
logarithmic region.

\subsection{All-order analysis}
\label{sec::3.2}

\begin{figure}[t]
\begin{center}
\begin{tabular}{ccc}
\includegraphics[width=3cm]{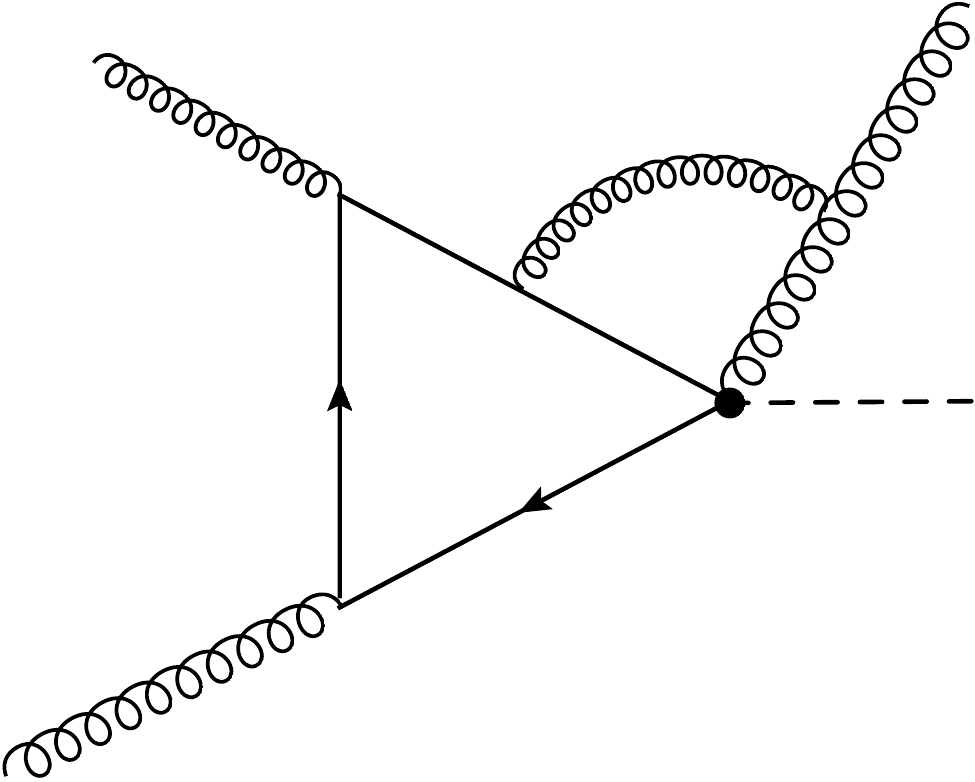}\hspace*{10.mm}&
\includegraphics[width=3cm]{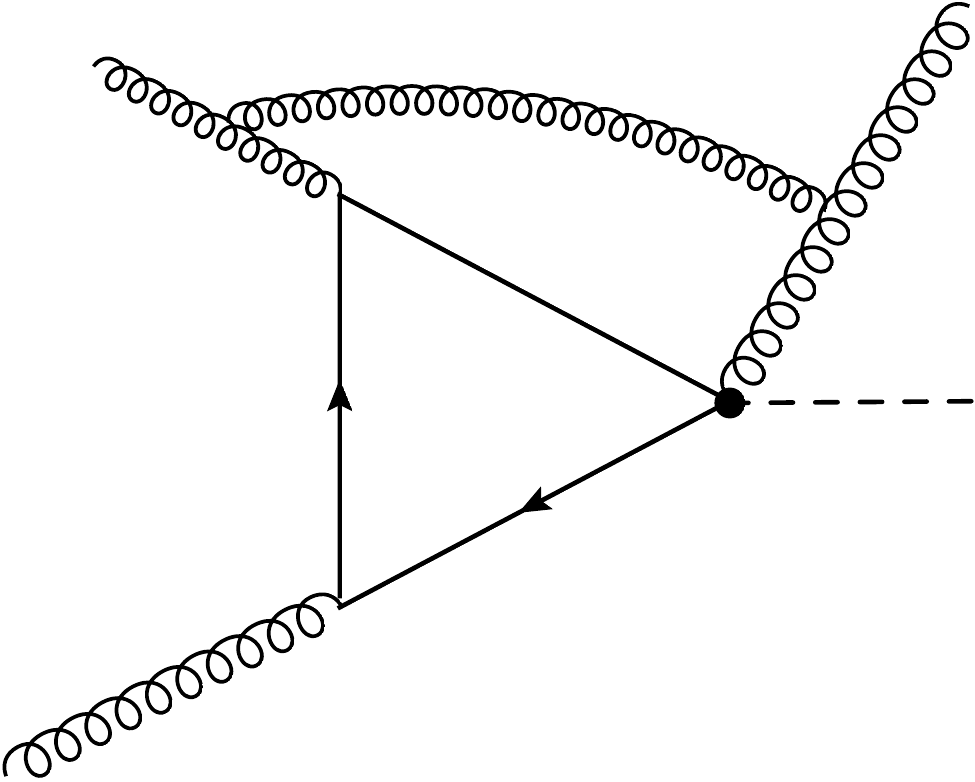}\hspace*{10.mm}&
\includegraphics[width=3cm]{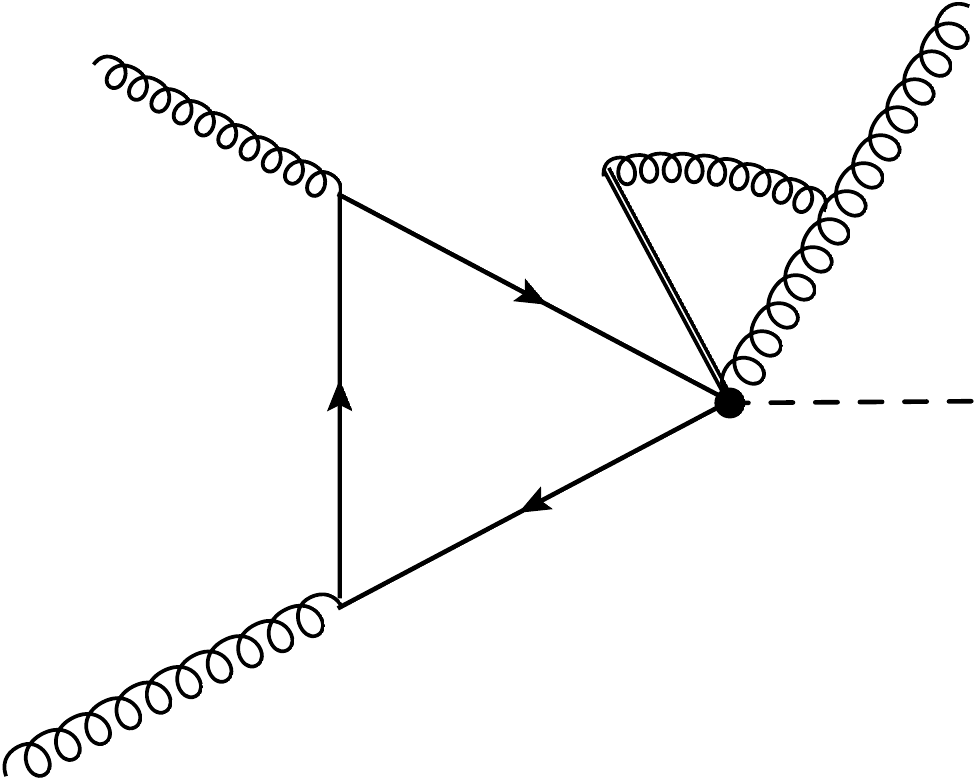}\\
&&\\
(a)&(b)&(c)\\
&&\\
\includegraphics[width=3cm]{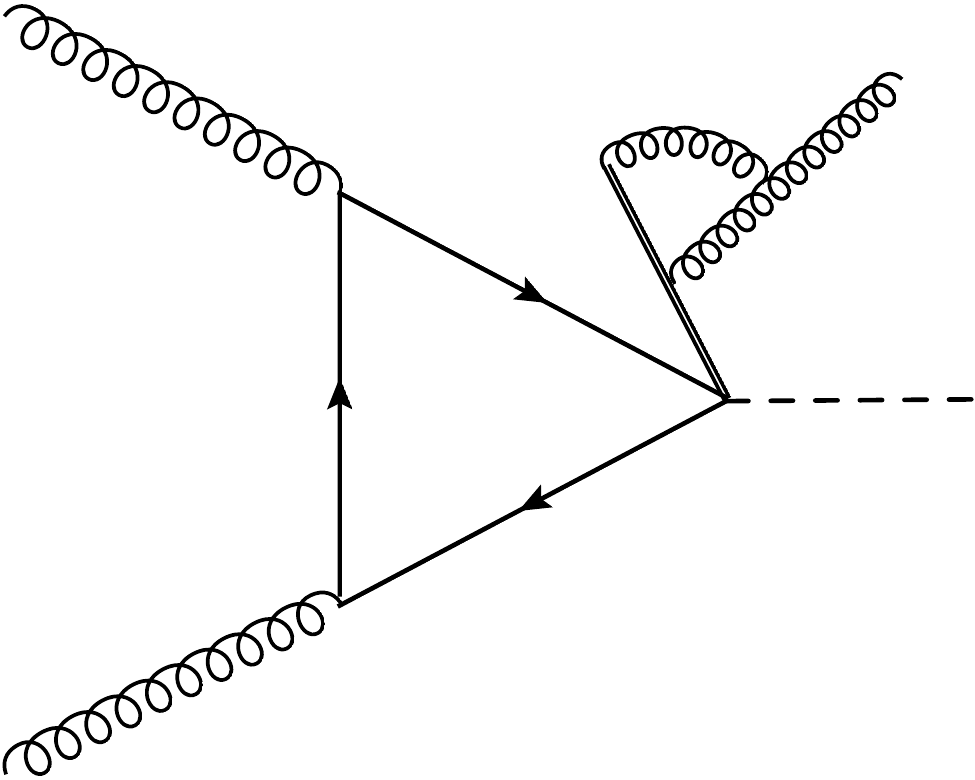}\hspace*{10.mm}&
\includegraphics[width=2.8cm]{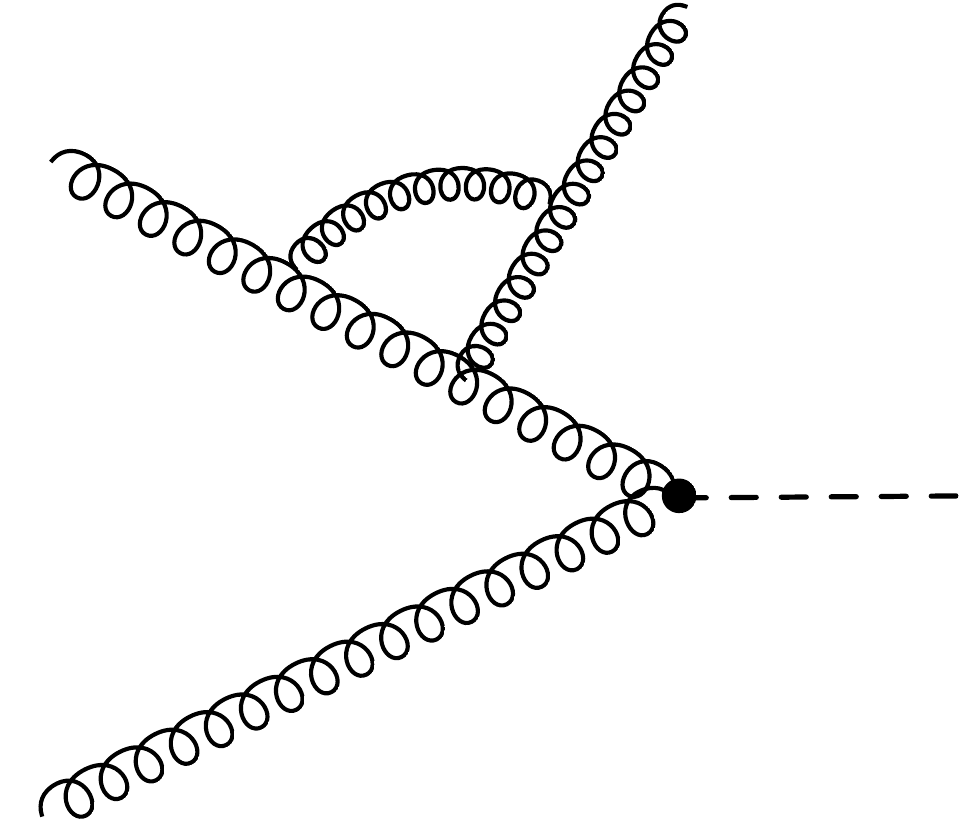}\hspace*{10.mm}&
\includegraphics[width=2.8cm]{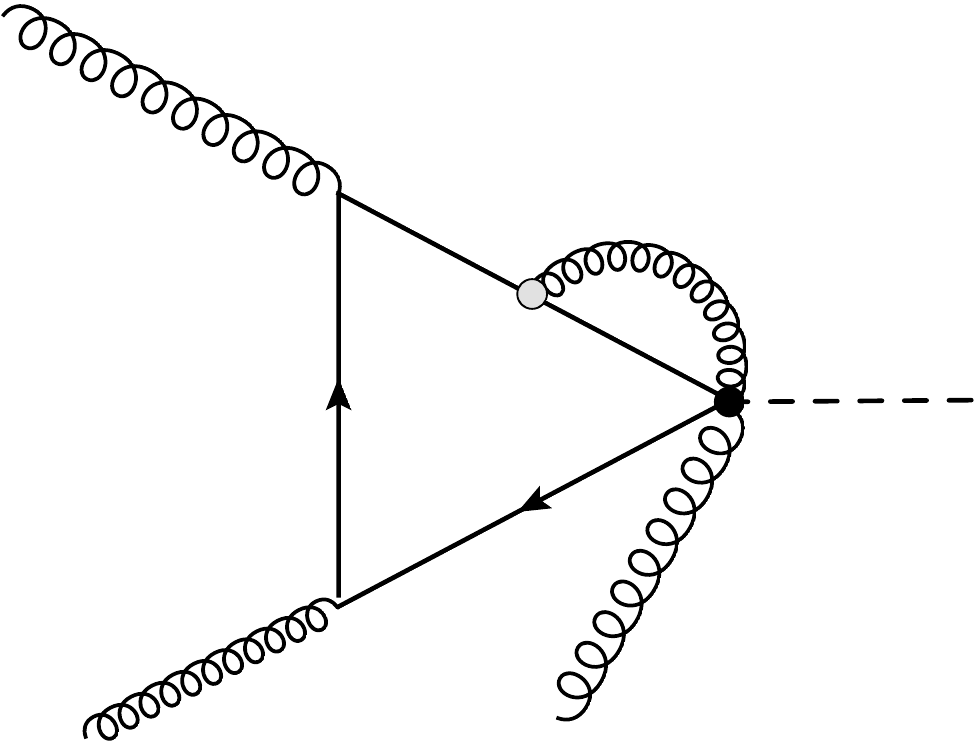}\hspace*{0.mm}\\
&&\\
(d)\hspace*{-20.mm}&(e)&(f)
\end{tabular}
\end{center}
\caption{\label{fig::4}  (a,b,c) Factorization of the soft
interaction of the initial state gluon in the symmetric
contribution. The eikonal double line carries the jet
momentum. (d) The factorized  soft interaction of the final
state gluon in the dipole contribution. The eikonal double line
carries the initial state momentum. (e) The heavy quark limit
diagram with the same structure of the  infrared Sudakov
logarithms as  the diagrams (c) and (d). (f) The Sudakov-Wilson
contribution with the soft gluon momentum  resolving the
collapsed propagator. The empty circle represents the dipole
interaction with the adjusted color weight.}
\end{figure}

The result of the previous section makes  the all-order
resummation of the double logarithmic radiative corrections
rather straightforward. These corrections are generated by the
soft virtual gluons with the well known factorization and
exponentiation properties \cite{Sudakov:1954sw,Frenkel:1976bj}
which should be adjusted to account for the non-Sudakov
logarithms in the soft quark initiated processes
\cite{Liu:2017vkm,Liu:2018czl}. The latter are characteristic
to the interaction of the eikonal lines connected by the soft
quark. Thus we do not expect the non-Sudakov contribution from
the interaction of the decoupled gluons to the quark loop. An
example of the eikonal factorization of this interaction  for
the symmetric contribution is shown in Fig.~\ref{fig::4}. If
the soft gluon momentum does not resolve the local vertex we
have
Fig.~\ref{fig::4}(a)$+$Fig.~\ref{fig::4}(b)$=$Fig.~\ref{fig::4}(c).
Note that the interaction of the virtual gluon with the soft
quark has the vanishing color factor. For the dipole
contribution similar factorization results in the diagram
Fig.~\ref{fig::4}(d). Both diagrams Fig.~\ref{fig::4}(c,d) have
the same structure of the Sudakov logarithms as the effective
theory diagram  Fig.~\ref{fig::4}(e) with the local
$g(p_1)g(p_2)H$ vertex.  Hence with our definition of the
operator $I^{(1)}_{\rm ph}$ all the corresponding  singular and
finite double logarithmic terms are included in $Z_{3g}$ factor
and we are left with the soft corrections to a quark loop
mediated $ggH$ amplitude, where one of the gluon can be the
jet. The structure of the double-logarithmic
corrections to such an amplitude is well understood too. In
general, after factoring out  Sudakov corrections absorbed into
$Z_{3g}$ we get the non-Sudakov double logarithmic contribution
associated with $A_{++\pm}$.  The structure of the
factorization and the non-Sudakov logarithms, however, is
different for the dipole and symmetric contributions.  Let us
first discuss the eikonal dipole part of the result. In this
case the jet completely decouples from the quark loop and one
can use the result \cite{Liu:2017vkm,Liu:2018czl} for the
$g(p_1)g(p_2)H$ amplitude, which gives
\begin{equation}
\left[A_{++\pm}^{\rm LL}\right]_{\rm e.d.} =
\pm 2L^2 \int_{0}^{1}{\rm d}\eta
\int_0^{1-\eta}{\rm d}\xi \,e^{2z\eta\xi}
=\pm L^2 \int_{0}^{1}{{\rm d}\eta\over z\eta }
\left(e^{2z\eta(1-\eta)}-1\right)
=\pm L^2g(z)\,,
\label{eq::eikLL}
\end{equation}
where $g(z)={}_2F_2\left(1,1;{3/2},2;{z/2}\right)$ is the
hypergeometric function, $z=(C_A-C_F)x$, $C_F={N_c^2-1\over
2N_c}$, and  $x = \frac{\alpha_s(\mu_s) }{4\pi}L^2$. The
exponent $e^{2z\eta\xi}$ in  Eq.~(\ref{eq::eikLL}) is the
off-shell $\bar qqH$ Sudakov form factor with the color weight
$C_F$ replaced by $C_F-C_A$ accounting for  the variation of
the color group representation for a particle moving along the
eikonal lines from adjoint to fundamental in the process of the
soft quark emission, which is the physical origin of the
non-Sudakov logarithms.

The soft dipole contribution differs from the eikonal one by
the  cutoff of the double-logarithmic region of virtual
momentum. This is an  {\it ultraviolet} cutoff which does not
change the infrared structure of the factorization and
therefore can be imposed directly in Eq.~(\ref{eq::eikLL}).
This gives
\begin{equation}
\left[A_{++\pm}^{\rm LL}\right]_{\rm s.d.} =
\mp L^2 \int_{\tau_t-\tau}^{\tau_t}{\rm d}\eta
\int_{1-\tau_t}^{1-\eta}{\rm d}\xi\,e^{2z\eta\xi}
=\mp L^2 \int_{\tau_t-\tau}^{\tau_t}
{{\rm d}\eta\over 2z\eta}
\left(e^{2z\eta(1-\eta)}
-e^{2z\eta(1-\tau_t)}\right)\,,
\label{eq::softdipLL}
\end{equation}
{\it cf.} Eqs.~(\ref{eq::softdip},\,\ref{eq::reseik}).

For the symmetric  contribution the eikonal lines carry the
momenta $p_3$ and $p_1$ or $p_2$. Thus  for {\it i.e.}
$g(p_1)g(p_3)H$ amplitude the exponential factor of
Eq.~(\ref{eq::eikLL}) should be changed to
$e^{2z(\eta-\tau_t+\tau)(\xi-1+\tau_t)}$, while the integration
limits are the same as in Eq.~(\ref{eq::softdipLL}).  However,
if a light-cone component of the virtual soft gluon momentum
exceeds the one of $p_3$, it may  resolve the local effective
vertex and such a correction is not taken into account by the
above exponent.  In this case the jet becomes soft with respect
to the virtual gluon momentum  and factors out of the
corresponding loop integral. Thus, the factorization of the
Sudakov corrections is realized  in the same way as for the
$g(p_1)g(p_2)H$ amplitude. The resulting non-Sudakov
contribution is accounted for by introducing a
``Sudakov-Wilson'' factor $e^{2z(\tau_t-\tau)\xi}$ local along
one of the eikonal lines, Fig.~\ref{fig::4}(f), with the
adjusted $C_F-C_A$ color weight. Compared to
Eq.~(\ref{eq::eikLL}), in this expression the $\tau_t-\tau$
factor replaces the logarithmic variable $\eta$ as the infrared
cutoff for the collapsed  quark propagator in the effective
vertex. Combining  the corrections we get
\begin{eqnarray}
\left[A_{++-}^{\rm LL}\right]_{\rm sym.}&=&
-L^2 \left[\int_{\tau_t-\tau}^{\tau_t}{\rm d}\eta
\int_{1-\tau_t}^{1-\eta}{\rm d}\xi\,
e^{2z(\eta-\tau_t+\tau)
(\xi-1+\tau_t)}e^{2z(\tau_t-\tau)\xi}
+(\tau_t\to\tau_u)\right]
\nonumber\\
&=&- L^2\left[ \int_{\tau_t-\tau}^{\tau_t}
{{\rm d}\eta\over 2z\eta}
e^{2z(\tau_t-\tau)(1-\tau_t)}
\left(e^{2z\eta(\tau_t-\eta)}-1\right)
+(\tau_t\to\tau_u)\right].
\label{eq::symLL}
\end{eqnarray}
Thus the effective theory decomposition of the
symmetric contribution has a  more complex structure
with the short-distance effects  described by the
Sudakov-Wilson factor.

We have verified the above factorization structure of
the amplitudes by explicit identification and
evaluation of the double-logarithmic contributions to the
two-loop diagrams using the technique  \cite{Liu:2018czl}.

The total  all-order leading logarithmic approximation
for the helicity form factors can be written in a compact
form
\begin{eqnarray}
A_{+++}^{LL} &=&L^2\bigg[g(z)-
\int_{(1-\tau+\zeta)/2}^{(1+\tau+\zeta)/2}
{{\rm d}\eta\over 2z\eta}
\left(e^{2z\eta(1-\eta)}
-e^{z\eta(1-\tau-\zeta))}\right)\bigg],
\label{eq::fullLLp}\\
A_{++-}^{LL} &=&-A_{+++}^{LL}-
L^2\bigg[
\int_{(1-\tau+\zeta)/2}^{(1+\tau+\zeta)/2}
{{\rm d}\eta\over 2z\eta}\,e^{z((1-\tau)^2-\zeta^2)/2}
\left(e^{z\eta(1+\tau+\zeta-2\eta)}-1\right)
+(\zeta\to-\zeta)\bigg],
\nonumber\\
&& \label{eq::fullLLm}
\end{eqnarray}
where $\zeta=\ln(t/u)/L$ is the rapidity variable. At the
boundary values of the allowed transverse momentum interval
Eqs.~(\ref{eq::fullLLp},\,\ref{eq::fullLLm}) simplify and can
be obtained in a closed analytic form. For $p_\perp\to m_b$
corresponding to $\tau\to 0$ the soft dipole and symmetric
contributions vanish and  we get
\begin{equation}
A_{++\pm}^{LL}|_{\tau=0}=\pm L^2g(z)\,.
\label{eq::tau0}
\end{equation}
At the same time for central rapidity $\zeta=0$ and
$p_\perp\to m_H$ corresponding to $\tau\to 1$ we have
$|s|\sim|t|\sim|u|$  and up to an overall factor
the symmetric, soft and eikonal dipole   contributions
are determined  by the same $ggH$ form factor. This gives
\begin{equation}
A_{+++}^{LL}|_{\tau=1,\,\zeta=0}={1\over 2}L^2{g(z)}\,, \qquad
A_{++-}^{LL}|_{\tau=1,\,\zeta=0}=-{3\over 2}L^2{g(z)}\,.
\label{eq::tau1}
\end{equation}
By expanding Eqs.~(\ref{eq::fullLLp},\,\ref{eq::fullLLm})
in $z$ we get the two-loop amplitudes
\begin{eqnarray}
A_{+++}^{(2)}&=&{(C_A-C_F)L^4\over 24}
\left(2-{3\tau^2}+{2\tau^3}+{3\tau^2\zeta^2}\right),
\label{eq::A2loopp}\\
A_{++-}^{(2)} &=&-{(C_A-C_F)L^4\over 24}
\left(2+{3\tau^2}-{6\tau^3}
+{4\tau^4}-{3\tau^2\zeta^2}\right),
\label{eq::A2loopm}
\end{eqnarray}
which can be compared to the result of the explicit
calculation \cite{Melnikov:2016qoc}. The comparison, however,
is not straightforward since in \cite{Melnikov:2016qoc} the
canonical symmetric  form of the Catani's operator has been
adopted
\begin{equation}
I^{(1)}_{\rm sym}=-{C_A\over 2\epsilon^2}
\left[\left({-s\over \mu^2}\right)^{-\epsilon}
+\left({-t\over \mu^2}\right)^{-\epsilon}
+\left({-u\over \mu^2}\right)^{-\epsilon}\right],
\label{eq::I1sym}
\end{equation}
which differs from Eq.~(\ref{eq::I1ph}) by
\begin{equation}
\delta I^{(1)}=-{C_A\over 8}L^2
\left[\left(1-\tau\right)^2-\zeta^2\right].
\label{eq::delI1}
\end{equation}
Thus for the helicity form factors defined in
\cite{Melnikov:2016qoc}  we need to perform the infrared
matching \cite{Penin:2005kf,Penin:2005eh}
\begin{equation}
\bar\Omega_{+++}^{(2l),{\rm fin}}=
A_{+++}^{(2)}+\delta I^{(1)} A_{+++}^{(1)}\,, \qquad
-\bar\Omega_{++-}^{(2l),{\rm fin}}=
A_{++-}^{(2)}+\delta I^{(1)}A_{++-}^{(1)}\,,
\label{eq::matching}
\end{equation}
which gives
\begin{eqnarray}
\bar\Omega_{+++}^{(2l),{\rm fin}} &=&-{L^4\over 24}
\left[C_A+2C_F-6C_A\tau+\left({9\over 2}C_A-3C_F\right){\tau^2}
+\left(C_A+2C_F\right){\tau^3}\right.
\nonumber\\
&&\left.-3C_A\zeta^2
+\left(-{3\over 2}C_A+3C_F\right)\tau^2\zeta^2\right],
\label{eq::Omega2loopp}
\\
\bar\Omega_{++-}^{(2l),{\rm fin}} &=&-{L^4\over 24}
\left[C_A+2C_F-6C_A\tau
+\left({3\over 2}C_A+3C_F\right){\tau^2}
+\left(3C_A+6C_F\right){\tau^3}\right.
\nonumber\\
&&\left.+\left(-{5\over 2}C_A+4C_F\right)\tau^4-3C_A\zeta^2
+\left({3\over 2}C_A-3C_F\right)\tau^2\zeta^2\right].
\label{eq::Omega2loopm}
\end{eqnarray}
Upon changing the notations $\zeta\to\xi$, $\tau\to-\tau$, for
$C_A=3$, $C_F=4/3$,
Eqs.~(\ref{eq::Omega2loopp},\,\ref{eq::Omega2loopm}) agree
with Eqs.~(6.11,\,6.12) of \cite{Melnikov:2016qoc}, providing
a nontrivial check of our analysis. Note that for the symmetric
choice of the infrared subtraction the physical $C_F-C_A$ color
scaling is missing.

\section{Logarithmic corrections to the kinematical distributions}
\label{sec::4}
We now can apply the result of the previous section to estimate
the bottom quark effect  to the differential cross section of
the Higgs boson production in association with a jet. The
dominant  contribution is due to the interference of the top
and bottom quark mediated amplitudes. We discard the partonic
channel with the initial state quarks and consider  the
numerically dominant contribution of the gluon fusion.  In the
physical inclusive cross section the processes with additional
real emission should be taken into account. If the  real
emission energy $E_{\rm real}$ exceeds $m_b$, it resolves the
bottom quark loop and may generate a new type of the double
logarithms. We, however, treat all the real emission as in the
heavy quark limit with the local $ggH$ interaction. This
approximation is discussed at the end of the section. Then both
the virtual Sudakov and the soft real corrections are the same
in the bottom and  top quark mediated processes and the
correction to the differential partonic cross section can be
written as follows
\begin{equation}
d\sigma^{tb}_{gg\to Hg+X}
=-\frac{3 m_b^2}{m_H^2} L^2 C_tC_b(\tau,\zeta)
d\tilde\sigma^{\rm eff}_{gg\to Hg+X}\,.
\label{eq::sigmapart}
\end{equation}
where $d\tilde\sigma^{\rm eff}_{gg\to Hg+X}$ is computed in the heavy
top effective theory and
\begin{eqnarray}
C_b(\tau,\zeta)&=&{A_{+++}-A_{++-}\over 2L^2}=
1+{z \over 6}\left(1-\tau^3+\tau^4\right) \nonumber \\
&+&{z^2}\left[{1\over 45}-{\tau^3\over 12}+{\tau^4\over 6}-
{7\tau^5\over 60}+{\tau^6\over 30} +{\zeta^2\over 12}
(\tau^3-\tau^4)\right]
\nonumber\\
&+&
{z^3}\left[{1\over 420}-{\tau^3\over 48}+{\tau^4\over 16}-
{\tau^5\over 12}+{23\tau^6\over 360}-
{143\tau^7\over 5040}+{31\tau^8\over 5040}\right.
\nonumber\\
&+&
\left.{\zeta^2}\left({\tau^3\over 24}
-{\tau^4\over 12}+{\tau^5\over 20}
-{\tau^6\over 180}\right)
-{\zeta^4\over 48}(\tau^3-\tau^4)\right]
+\ldots
\label{eq::Cbtz}
\end{eqnarray}
is a function of the transverse momentum and rapidity.
Convolution of Eq.~(\ref{eq::sigmapart})  with the parton
distribution functions gives the correction to the kinematical
distributions of the $pp\to Hj+X$ production in the threshold
approximation.

\begin{figure}[t]
\begin{center}
\begin{tabular}{c}
\includegraphics[width=8cm]{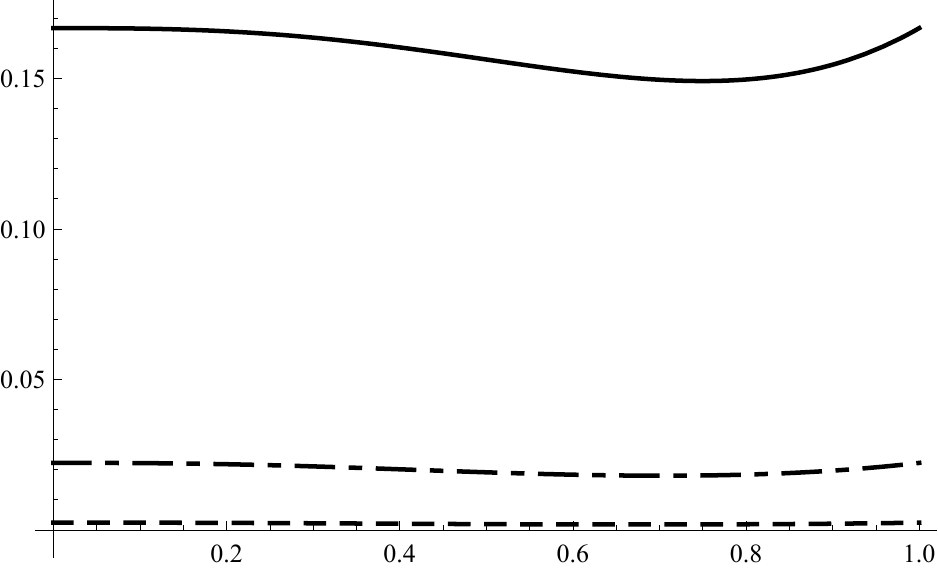}\\
$\tau$
\end{tabular}
\end{center}
\caption{\label{fig::5}
The transverse momentum dependence of  $z^n$  coefficients in
Eq.~(\ref{eq::Cbtz}) at central rapidity $\zeta=0$ for $n=1$
(solid line), $n=2$ (dot-dashed line) and $n=3$ (dashed line).
For each coefficient the boundary values at $\tau=0,~1$ are
equal.}
\end{figure}

However, within a well-motivated approximation the analysis of
the hadronic cross section can be significantly simplified
\cite{Melnikov:2016emg}. Indeed, the dependence of
Eq.~(\ref{eq::Cbtz}) on the jet rapidity is very weak. It
starts with ${\cal O}(z^2)$ and the rapidity-dependent terms
include at  least the second power of the variable $\zeta$ and
the third power of the variable $\tau$.  If  the soft jet is
emitted at large rapidity, we have $|\zeta| \approx 1$ and
$\tau\ll 1$. On the contrary, central emission with the large
transverse momentum implies $|\zeta| \ll 1$ and $\tau \approx
1$. Therefore, the rapidity-dependent terms are small
everywhere  and can be neglected.  Then $C_b(\tau,0)$ factors
out from the parton distribution function integral. At the same
time the transverse momentum dependence of the coefficients in
Eq.~(\ref{eq::Cbtz}) turned out to be very weak too,
Fig.~\ref{fig::5}, and it can be approximated by $C_b=
C_b(0,\zeta)=C_b(1,0)$ with a very high accuracy. This is quite
a nontrivilal fact since the individual helicity form factors
strongly depend on the transverse momentum,
Eqs.~(\ref{eq::tau0},\,\ref{eq::tau1}). The coefficient
$C_b=1+\sum_{n=1}^\infty c_n$  is saturated by the eikonal
dipole part of the factorization formula with the jet emission
decoupled from the quark loop and  describes the non-Sudakov
logarithmic corrections to the total cross section of the Higgs
boson production.  It is known through the NLL approximation
\cite{Anastasiou:2020vkr}. The first three terms of the series
read
\begin{eqnarray}
c_1&=&{z\over 6}+C_F{\alpha_s L\over 4\pi}\,,
\nonumber\\
c_2&=&{z^2\over 45}+{x\over 5}{\alpha_s L\over 4\pi}
\left[{3\over 2}C_F-\beta_0\left({5\over 6}{L_\mu\over L}
-{1\over 3}\right)\right],
\label{eq::Cb}\\
c_3&=&{z^3\over 420}+{x^2\over 5}{\alpha_s L\over 4\pi}
\left[{5\over 21}C_F-\beta_0\left({2\over 9}{L_\mu\over L}
-{2\over 21}\right)\right],
\nonumber
\end{eqnarray}
where $\beta_0={11\over 3}C_A-{4\over 3}T_Fn_l$, $n_l=5$,
$L_\mu=\ln(s/\mu^2)$ and $\mu$ is the renormalization scale of
$\alpha_s$ in the double-logarithmic variable $z$. Thus
neglecting the tiny soft dipole and symmetric contributions to
the factorization formula we can extend the NLL analysis of the
total threshold cross section  to the process with the final
jet. This gives
\begin{equation}
{d\sigma^{tb}_{pp\to Hj+X}}=-\frac{3 m_b^2}{m_H^2}
\left({\alpha_s(m_H)\over \alpha_s(m_b)}\right)^{\gamma_m^{(1)}/\beta_0}
L^2 {C_b C_t}{d\tilde\sigma^{\rm eff}_{pp\to Hj+X}}\,,
\label{eq::sigmahad}
\end{equation}
where $\gamma^{(1)}_m={3}C_F$  is the one-loop mass anomalous
dimension and the renormalization group factor sets the
physical scale of the bottom quark Yukawa coupling to $m_H$.
Clearly, in this approximation we miss  the  NLL terms in
$C_b(\tau,\zeta)$ with the non-trivial dependence on the
kinematical variables. However, the analysis of the total cross
section  \cite{Anastasiou:2020vkr} indicates that the main
effect of the non-Sudakov NLL contribution is in fixing the
physical renormalization scales of the strong and Yukawa
couplings in the leading-logarithmic result, which is included
into  Eq.~(\ref{eq::sigmahad}).

It is instructive to rewrite this formula in a way which
relates the $K$-factors for the  top-bottom interference
contribution and the top quark mediated cross section in the
heavy top limit
\begin{equation}
{d\sigma^{tb}_{pp\to Hj+X}}=
\left[{C_b\over  C_t}
\left({\alpha_s(m_H)\over \alpha_s(m_b)}
\right)^{\gamma_m^{(1)}/\beta_0} \right]
\left({d\sigma^{tb}_{pp\to Hj+X}\over
d\sigma^{tt}_{pp\to Hj+X}}\right)^{\rm LO}
{d\sigma^{tt}_{pp\to Hj+X}}\,.
\label{eq::sigmarel}
\end{equation}
Here the first factor accounts for  the difference between the
QCD corrections to the top and bottom quark contributions. In
the above approximation this difference does not depend on the
kinematical variables and does not exceed twenty percent in the
NLO, in agreement with  the full calculation
\cite{Lindert:2017pky}. The NNLO term in the expansion of  this
factor amounts of  a few percent so that the total NNLO
correction to the bottom quark contribution is dominated  by
the K-factor  of  the top quark mediated cross section
\cite{Chen:2014gva,Boughezal:2015dra,Boughezal:2015aha,Caola:2015wna,Chen:2016zka}.
Note that in the NLL approximation there is no difference
between the pole mass $m_b$ and the $\overline{\rm MS}$ mass
$\bar{m}_b(\bar{m}_b)$ but the use of  the latter as the bottom
quark mass parameter in the available exact fixed order results
\cite{Anastasiou:2020vkr} boosts the convergence of the
perturbative expansion ({\it cf.} \cite{Czakon:2023kqm}).

An important property of our result is that at $p_\perp\to m_b$
the soft dipole and symmetric  parts of the factorization
formula vanish and in the remaining eikonal part the jet is
completely decoupled from the quark loop. Thus, the transverse
momentum evolution of the cross section matches the one for
$p_\perp\ll m_b$ derived in the heavy bottom  effective theory
with $C_b$ playing the role of the Wilson coefficient of the
local $ggH$ interaction. Moreover, up to a tiny corrections
even for $m_q\ll p_\perp$  the jet emission factors out as in
the large quark mass limit. We may assume a similar behavior  of
the unresolved real emission with the energy below the  energy
of the jet. Then our result, being formally valid
for $E_{\rm real}\ll m_b$, can be extended to the entire energy
region where the real emission is not kinematically suppressed.
Such a behavior is supported by the observed structure of the
NLO corrections \cite{Lindert:2017pky} and explains the accuracy of
the NNLO result for the  total cross section
\cite{Anastasiou:2020vkr}, where  the (soft) real emission
was approximated by the  heavy quark limit expression.

Thus our result can be used to complete the renormalization
group analysis of the bottom quark contribution to the
transverse momentum observables. The well-understood
logarithms of the ratio $p_\perp/m_H$ in the  effective theory
cross section $d\tilde\sigma^{\rm eff}_{pp\to Hj+X}$ are known
to very high orders of the logarithmic expansion
\cite{Banfi:2015pju, Bizon:2017rah}. At the same time the
non-Sudakov logarithms discussed above are missing in the
existing analysis of the bottom quark effects
\cite{Banfi:2013eda,Caola:2018zye}. They can be fully taken
into account  in the  leading-logarithmic approximation through
Eq.~(\ref{eq::sigmapart}). However, as it has been explained
above, their effect can be very well approximated by
Eq.~(\ref{eq::sigmahad}) with the all-order NLL expression for
$C_b$ available in \cite{Anastasiou:2020vkr}. Note that the
multiple soft  emission with the transverse momenta of order
$p_\perp$ can in general resolve the bottom quark loop and the
corresponding logarithmic corrections  differ from the
effective theory result. However, the arguments  given above
indicate that the difference is numerically small. Moreover,
with the increasing number of soft real partons their energies
in the kinematically unsuppressed   region  are reduced and
the heavy bottom  effective theory becomes more and more
accurate.

\section{Conclusion}
\label{sec::conc}
In this paper, we have studied a  light quark mediated
production of the Higgs boson  in association with a jet in
gluon fusion in the double-logarithmic approximation for  the
intermediate values of transverse momentum $m_q\ll p_\perp\ll
m_H$. The factorization formula has been obtained for the
helicity amplitudes, which separates the universal infrared
Sudakov renormalization of the external on-shell gluon states
and the non-Sudakov logarithms characteristic to the power
suppressed contributions. The physical origin of the latter is
the color charge nonconservation  due to exchange of a soft
quark between  each pair of the eikonal lines associated with
the external on-shell gluons, as for the $gg\to H$ amplitude.
Thus, despite significantly more complex geometry of the
double-logarithmic region, with the physical definition of the
infrared subtraction operator the full QCD result can be
obtained from its abelian part by the color factor adjustment
$C_F\to C_F-C_A$.   The non-Sudakov double-logarithmic
corrections to the $gg\to Hg$ partonic cross section  show very
weak dependence on the jet transverse momentum and rapidity  in
the whole allowed region of the kinematical variables and for
the both boundary values $p_\perp\to m_q,\, m_H$   reduce
precisely to the  $gg\to H$ case.\footnote{This is not valid
for the individual helicity amplitudes at $p_\perp\to m_H$.}

This indicates that at least in the leading logarithmic
approximation up to a tiny corrections the jet emission factors
out even for $m_q\ll p_\perp$ as in the large quark mass limit
but with a different mass dependence of the Wilson coefficient.
This, in particular, explains the observed structure of the NLO
result \cite{Lindert:2017pky} as well as the success of the
NNLO  analysis of the total cross section
\cite{Anastasiou:2020vkr}, with the real emission  being
treated in the same way as for the top quark mediated process.

Relying on this property we have  derived  the NNLO bottom
quark contribution to the kinematical distributions of the
$pp\to Hj+X$ production with the logarithmic accuracy in terms
of the transverse momentum dependent $K$-factors previously
computed in the  heavy top quark limit
\cite{Chen:2014gva,Boughezal:2015dra,Boughezal:2015aha,Caola:2015wna,Chen:2016zka}.
In fact, the NNLO $K$-factors for the top-bottom interference
and the top quark contribution to the kinematical distributions
should agree within a few percent in the given region of
transverse momentum.

Our all-order result for the  amplitudes also  provides the
last missing ingredient of the renormalization group analysis
of the bottom quark effect in the Higgs boson plus jet
production at the LHC \cite{Banfi:2013eda,Caola:2018zye}. To
our knowledge this is the first double-logarithmic asymptotic
result for a  QCD amplitude which captures all-order dependence
on two kinematical variables, {\it i.e.} simultaneously sums up
three different types of the large logarithms.

\section*{Acknowledgments}
A.P is grateful to Babis Anastasiou and Kirill Melnikov for
many helpful discussions and comments. A.R. would like to thank
Lorenzo Tancredi for useful communications. The work of T.L. is
supported in part by IHEP under Grants No.~Y9515570U1, the
National Natural Science Foundation of China (NNSFC) under
grant No.12375082 and No.12135013.  The work of A.P. is
supported in part by NSERC and Perimeter Institute for
Theoretical Physics. The work of A.R. is supported by NSERC.


\begin{thebibliography}{99}


\bibitem{ATLAS:2016neq}
G.~Aad \textit{et al.} [ATLAS and CMS],
JHEP \textbf{08}, 045 (2016).


\bibitem{Cepeda:2019klc}
M.~Cepeda, S.~Gori, P.~Ilten, M.~Kado, F.~Riva, R.~Abdul Khalek, A.~Aboubrahim, J.~Alimena, S.~Alioli and A.~Alves, \textit{et al.}
CERN Yellow Rep. Monogr. \textbf{7}, 221 (2019).


\bibitem{Arnesen:2008fb}
C.~Arnesen, I.~Z.~Rothstein and J.~Zupan,
Phys. Rev. Lett. \textbf{103}, 151801 (2009).



\bibitem{Bagnaschi:2011tu}
E.~Bagnaschi, G.~Degrassi, P.~Slavich and A.~Vicini,
JHEP \textbf{02}, 088 (2012).


\bibitem{Dawson:2014ora}
S.~Dawson, I.~M.~Lewis and M.~Zeng,
Phys. Rev. D \textbf{90},  093007 (2014).


\bibitem{Grazzini:2016paz}
M.~Grazzini, A.~Ilnicka, M.~Spira and M.~Wiesemann,
JHEP \textbf{03}, 115 (2017).



\bibitem{Chen:2014gva}
X.~Chen, T.~Gehrmann, E.~W.~N.~Glover and M.~Jaquier,
Phys. Lett. B \textbf{740}, 147 (2015).

\bibitem{Boughezal:2015dra}
R.~Boughezal, F.~Caola, K.~Melnikov, F.~Petriello and M.~Schulze,
Phys. Rev. Lett. \textbf{115}, 082003 (2015).

\bibitem{Boughezal:2015aha}
R.~Boughezal, C.~Focke, W.~Giele, X.~Liu and F.~Petriello,
Phys. Lett. B \textbf{748}, 5 (2015).

\bibitem{Caola:2015wna}
F.~Caola, K.~Melnikov and M.~Schulze,
Phys. Rev. D \textbf{92}, 074032 (2015).

\bibitem{Chen:2016zka}
X.~Chen, J.~Cruz-Martinez, T.~Gehrmann, E.~W.~N.~Glover and M.~Jaquier,
JHEP \textbf{10}, 066 (2016).

\bibitem{Jones:2018hbb}
S.~P.~Jones, M.~Kerner and G.~Luisoni,
Phys. Rev. Lett. \textbf{120},  162001 (2018).


\bibitem{Baur:1989cm}
U.~Baur and E.~W.~N.~Glover,
Nucl. Phys. B \textbf{339}, 38-66 (1990).


\bibitem{Anastasiou:2016cez}
C.~Anastasiou, C.~Duhr, F.~Dulat, E.~Furlan, T.~Gehrmann, F.~Herzog, A.~Lazopoulos and B.~Mistlberger,
JHEP {\bf 1605}, 058 (2016).


\bibitem{Bishara:2016jga}
F.~Bishara, U.~Haisch, P.~F.~Monni and E.~Re,
Phys. Rev. Lett. \textbf{118},  121801 (2017).


\bibitem{Soreq:2016rae}
Y.~Soreq, H.~X.~Zhu and J.~Zupan,
JHEP \textbf{12}, 045 (2016).


\bibitem{Lindert:2017pky}
J.~M.~Lindert, K.~Melnikov, L.~Tancredi and C.~Wever,
Phys.\ Rev.\ Lett.\  {\bf 118},  252002 (2017).


\bibitem{Bonciani:2022jmb}
R.~Bonciani, V.~Del Duca, H.~Frellesvig, M.~Hidding, V.~Hirschi, F.~Moriello, G.~Salvatori, G.~Somogyi and F.~Tramontano,
Phys. Lett. B \textbf{843}, 137995 (2023).


\bibitem{Mantler:2012bj}
H.~Mantler and M.~Wiesemann,
Eur. Phys. J. C \textbf{73},  2467 (2013).

\bibitem{Grazzini:2013mca}
M.~Grazzini and H.~Sargsyan,
JHEP \textbf{09}, 129 (2013).


\bibitem{Banfi:2013eda}
A.~Banfi, P.~F.~Monni and G.~Zanderighi,
JHEP \textbf{01}, 097 (2014).

\bibitem{Caola:2018zye}
F.~Caola, J.~M.~Lindert, K.~Melnikov, P.~F.~Monni, L.~Tancredi and C.~Wever,
JHEP \textbf{09}, 035 (2018).

\bibitem{Kotsky:1997rq}
M.~I.~Kotsky and O.~I.~Yakovlev,
Phys.\ Lett.\ B {\bf 418}, 335 (1998).


\bibitem{Penin:2014msa}
A.~A.~Penin,
Phys.\ Lett.\ B {\bf 745}, 69 (2015).


\bibitem{Liu:2017axv}
T.~Liu, A.~A.~Penin and N.~Zerf,
Phys.\ Lett.\ B {\bf 771}, 492 (2017).

\bibitem{Sudakov:1954sw}
V.~V.~Sudakov,
Sov.\ Phys.\ JETP {\bf 3}, 65 (1956)
[Zh.\ Eksp.\ Teor.\ Fiz.\  {\bf 30}, 87 (1956)].


\bibitem{Frenkel:1976bj}
J.~Frenkel and J.~C.~Taylor,
Nucl.\ Phys.\ B {\bf 116}, 185 (1976).


\bibitem{Mueller:1979ih}
A.~H.~Mueller,
Phys.\ Rev.\ D {\bf 20}, 2037 (1979).

\bibitem{Collins:1980ih}
J.~C.~Collins,
Phys.\ Rev.\ D {\bf 22}, 1478 (1980).

\bibitem{Sen:1981sd}
A.~Sen,
Phys.\ Rev.\ D {\bf 24}, 3281 (1981).

\bibitem{Sterman:1986aj}
G.~F.~Sterman,
Nucl.\ Phys.\ B {\bf 281}, 310 (1987).

\bibitem{Liu:2017vkm}
T.~Liu and A.~A.~Penin,
Phys.\ Rev.\ Lett.\  {\bf 119},  262001 (2017).


\bibitem{Liu:2018czl}
T.~Liu and A.~Penin,
JHEP {\bf 1811}, 158 (2018).



\bibitem{Liu:2021chn}
T.~Liu, S.~Modi and A.~A.~Penin,
JHEP \textbf{02}, 170 (2022).

\bibitem{Melnikov:2016emg}
K.~Melnikov and A.~Penin,
JHEP {\bf 1605}, 172 (2016).

\bibitem{Penin:2016wiw}
A.~A.~Penin and N.~Zerf,
Phys.\ Lett.\ B {\bf 760}, 816 (2016).

\bibitem{Delto:2023kqv}
M.~Delto, C.~Duhr, L.~Tancredi and Y.~J.~Zhu,
[arXiv:2311.06385 [hep-ph]].

\bibitem{Penin:2019xql}
A.~A.~Penin,
JHEP {\bf 2004}, 156 (2020).

\bibitem{Anastasiou:2020vkr}
C.~Anastasiou and A.~Penin,
JHEP {\bf 2007}, 195 (2020).

\bibitem{Liu:2020wbn}
Z.~L.~Liu, B.~Mecaj, M.~Neubert and X.~Wang,
JHEP {\bf 2101}, 077 (2021).

\bibitem{Liu:2022ajh}
Z.~L.~Liu, M.~Neubert, M.~Schnubel and X.~Wang,
JHEP \textbf{06}, 183 (2023).


\bibitem{Czakon:2023kqm}
M.~Czakon, F.~Eschment, M.~Niggetiedt, R.~Poncelet and T.~Schellenberger,
[arXiv:2312.09896 [hep-ph]].


\bibitem{Schroder:2005hy}
Y.~Schroder and M.~Steinhauser,
JHEP {\bf 0601}, 051 (2006).

\bibitem{Chetyrkin:2005ia}
K.~G.~Chetyrkin, J.~H.~Kuhn and C.~Sturm,
Nucl.\ Phys.\ B {\bf 744}, 121 (2006).


\bibitem{Melnikov:2016qoc}
K.~Melnikov, L.~Tancredi and C.~Wever,
JHEP {\bf 1611}, 104 (2016).

\bibitem{Gehrmann:2011aa}
T.~Gehrmann, M.~Jaquier, E.~W.~N.~Glover and A.~Koukoutsakis,
JHEP \textbf{02}, 056 (2012).



\bibitem{Catani:1998bh}
S.~Catani,
Phys.\ Lett.\ B {\bf 427}, 161 (1998).


\bibitem{Schmidt:1997wr}
C.~R.~Schmidt,
Phys. Lett. B \textbf{413}, 391-395 (1997).


\bibitem{Penin:2005kf}
A.~A.~Penin,
Phys.\ Rev.\ Lett.\  {\bf 95}, 010408 (2005).

\bibitem{Penin:2005eh}
A.~A.~Penin,
Nucl.\ Phys.\ B {\bf 734},  185 (2006).

\bibitem{Banfi:2015pju}
A.~Banfi, F.~Caola, F.~A.~Dreyer, P.~F.~Monni, G.~P.~Salam, G.~Zanderighi and F.~Dulat,
JHEP \textbf{04}, 049 (2016).


\bibitem{Bizon:2017rah}
W.~Bizon, P.~F.~Monni, E.~Re, L.~Rottoli and P.~Torrielli,
JHEP \textbf{02}, 108 (2018).


\end{thebibliography}
\end{document}